# Direct Laser Micromachining of Superconducting Terahertz Josephson Plasma Emitters

## AUTHORS

Reo Yamaguchi,[1,2,a)] Takuma Sakurai,[1,2,a)] Kazuhiro Yamaki,[2] Akinobu Irie,[2] Junichiro Kato,[1,3,b)] Taichiro Nishio,[3] Shigeyuki Ishida,[1] Hiroshi Eisaki,[1] and Manabu Tsujimoto[1,2,c)]

## AFFILIATIONS

[1.] Core Electronics Technology Research Institute, National Institute of Advanced Industrial Science and Technology (AIST), Tsukuba, Ibaraki 305-8565/8568, Japan

[2.] Graduate School of Regional Development and Creativity, Utsunomiya University, Utsunomiya, Tochigi 321-8585, Japan

[3.] Department of Physics, Tokyo University of Science, Shinjuku, Tokyo 162-8601, Japan

[a)] These authors contributed equally to this work.
[b)] Present address: Institute of Particle and Nuclear Studies, High Energy Accelerator Research Organization (KEK), Tsukuba, Ibaraki 305-0801, Japan
[c)] Author to whom correspondence should be addressed: m.tsujimoto@aist.go.jp

## ABSTRACT

We demonstrate a rapid, maskless fabrication method for superconducting terahertz Josephson plasma emitters (JPEs) based on direct ultraviolet laser micromachining of $Bi_2Sr_2CaCu_2O_{8+\delta}$ (Bi-2212) single crystals. Although machining debris is formed near the processed regions, uniform stacks of intrinsic Josephson junctions are preserved inside the crystal, enabling stable terahertz emission. Devices fabricated with Ag, Cu, and Cr electrodes all exhibited terahertz radiation, with Cu electrodes showing performance comparable to Ag while offering a low-cost alternative. Spectroscopic and polarization analyses indicate that the emitted radiation is elliptically polarized and dominated by the geometrical cavity resonance mode. Structural and electrical characterizations reveal that the machining width and depth are not limited by the optical spot size but are governed by the anisotropic thermal conductivity of Bi-2212, consistent with a thermally dominated laser ablation process. This direct laser micromachining approach provides a fast and versatile fabrication technique for JPEs and is broadly applicable to superconducting electronics and terahertz devices.

Terahertz radiation, spanning frequencies from approximately 0.1 to 10 THz between the microwave and infrared regimes, is attracting broad interest for applications across condensed-matter physics, materials science, spectroscopy, medical diagnostics, security screening, and high-speed wireless communication.[1,2] In this frequency range, terahertz photons couple directly to low-energy excitations in solids, including phonons and spin excitations, as well as molecular rotational and vibrational transitions, enabling nondestructive and selective probing of matter.[3] A key challenge in terahertz science and technology is the development of compact, efficient, and electrically driven solid-state terahertz emitters. Although laboratory-scale sources based on large accelerator facilities or femtosecond laser systems provide high performance, their size, complexity, and cost hinder practical deployment. Chip-scale semiconductor emitters such as quantum cascade lasers[4,5] and resonant tunneling diodes[6] represent important advances, but their reliance on epitaxial growth and nanofabrication processes limits scalability and flexibility due to the need for cleanroom-based manufacturing.

Josephson plasma emitters (JPEs) based on layered cuprate superconductors offer a fundamentally different approach to solid-state terahertz generation.[7,8] In highly anisotropic cuprates such as $Bi_2Sr_2CaCu_2O_{8+\delta}$ (Bi-2212), intrinsic Josephson junctions (IJJs) are naturally formed by the atomic-scale stacking of superconducting and insulating layers,[9,10] enabling low-loss Josephson plasma oscillations over a broad frequency range from approximately 100 GHz to several THz.[11] Owing to the Josephson effect, the oscillation frequency can be continuously tuned by the applied bias voltage,[12–14] in contrast to semiconductor-based terahertz sources whose emission frequency is typically fixed by cavity dimensions or quantum-well structures. Moreover, the Bi-2212 mesa itself acts as a cavity resonator or planar antenna,[15–18] allowing control of emission frequency and polarization through geometric design. Recent demonstrations of terahertz emission from trilayer cuprate $Bi_2Sr_2Ca_2Cu_3O_{10+\delta}$[19] and broadband frequency-modulated oscillations[20] underscore the potential of JPEs for applications including terahertz imaging,[21,22] sensing, and wireless terahertz cryogenic interconnects.[23]

Despite these advantages, the practical deployment of JPEs is still constrained by the lack of simple and efficient fabrication methods. Conventional argon ion milling suffers from low throughput and sample heating, while focused ion beam (FIB) processing, although precise, is inefficient for micrometer-scale mesas and can introduce irradiation damage. Wet-etching approaches reduce processing complexity[24,25] but require lithographic masking, provide limited dimensional accuracy at tens-of-micrometers scales, and are inherently isotropic, restricting three-dimensional mesa design. Even recently proposed ultraviolet (UV)-assisted wet-etching techniques[26] essentially correspond to maskless exposure lithography rather than true direct micromachining.

In this study, we develop a rapid, thermally dominated direct laser micromachining technique for Bi-2212 single crystals that eliminates conventional lithographic and wet-etching processes. The high aspect-ratio mesa structures produced by this method exhibit steep and well-defined sidewalls, which are favorable for spontaneous phase synchronization of stacked IJJs and thus for efficient terahertz emission. In this work, we fabricate JPEs with three different electrode materials, including previously unexplored candidates, and systematically investigate their electrical characteristics and terahertz emission properties.

Single crystals of Bi-2212 were grown by the traveling solvent floating zone method. The carrier concentration was adjusted to an underdoped state by reduction annealing: the crystals were first annealed at 800 °C for 24 h in an oxygen atmosphere to homogenize the excess oxygen content, followed by annealing at 600 °C for 24 h in a nitrogen atmosphere and quenching to room temperature.

The fabrication process of the JPEs is schematically shown in Figs. 1(a-1)–1(a-3). In step (a-1), a thin Bi-2212 crystal flake was bonded onto a 7×7 mm$^2$ sapphire substrate using a thermally conductive paste and baked at 160 °C for 1 h. After cleavage, the sample was immediately transferred into a vacuum chamber, where metal electrodes were deposited by thermal evaporation at a base pressure below 5×10$^{-4}$ Pa.

The JPEs employ a bilayer electrode structure consisting of a bottom contact layer and an Au capping layer. In addition to the commonly used Ag electrodes, Cu and Cr were examined to evaluate the influence of electrode materials on contact resistance. Previous studies on YBa$_2$Cu$_3$O$_7$ (YBCO) superconductors have shown that electrode materials can affect interfacial properties through oxidation and oxygen deficiency,[27–29] but systematic comparisons for Bi-2212-based JPEs have been lacking. To minimize thermal and physical damage to the cleaved surface, metal deposition was initiated at a reduced rate of 0.6–0.8 Å/s until a thickness of 5 nm was reached. A 20-nm-thick Au capping layer was deposited for all samples to suppress oxidation. The device structures and electrode configurations are summarized in Table I.

In step (a-2), vertical trench structures with nominal widths of approximately 10 μm were directly patterned on the Bi-2212 surface using a UV laser marker. Compared with FIB trenching,[12,15] this optical micromachining method is limited in spatial resolution by the diffraction limit but offers a decisive advantage in processing speed, completing patterning within less than 1 s. A UV laser source with a wavelength of 355 nm was used for rapid two-dimensional patterning based on predefined computer-aided design (CAD) patterns.

Laser processing conditions were optimized in advance, as the machining characteristics are governed primarily by energy density and temporal energy delivery. Longer pulse widths enhance thermally dominated ablation via heat diffusion, while the repetition rate and scan speed control pulse overlap along the scan path. From systematic evaluation of trench morphology, depth, and sidewall profile, a pulse width of 5 μs, repetition rate of 80 kHz, and scan speed of 500 mm/s were identified as optimal and used throughout this study.

Representative scanning electron microscopy (SEM) images of the fabricated mesas are shown in Figs. 1(b-1) and 1(b-2). The CAD-based process enables flexible patterning of various mesa geometries and straightforward fabrication of mesa arrays on a single crystal (see Fig. 1(c)). Resolidified machining debris forms a characteristic caldera-like morphology around trench lines, with a thickness of approximately 10 μm. Although this debris prevents direct estimation of the internal machining depth by surface inspection, the effective trench depth can be reliably inferred from electrical measurements and emission frequency analysis, as discussed below.

In the final step (a-3), the top bias electrode was formed by wire bonding a 20-μm-diameter Au wire onto the isolated mesa, while the surrounding crystal base served as the ground electrode. The order of laser processing and wire bonding is interchangeable, allowing multistep fabrication of three-dimensional mesa structures. The fabricated JPE chips were mounted in a compact sample package using a phosphor-bronze clamp to ensure good thermal contact with the heat bath.[30] Subsequent electrical and terahertz emission measurements were carried out using a closed-cycle cryostat and standard bolometric detection.

The resistance–temperature ($R$–$T$) characteristics of Samples A, B, and C are shown in Fig. S1 of the Supplementary Material. As summarized in Table I, all samples exhibit similar superconducting transition temperatures of $T_c = 75$-$77$ K, indicating that the Bi-2212 single crystals are in an underdoped state compared with optimally doped Bi-2212. This doping level is determined primarily by the reduction-annealing treatment and is largely preserved during UV laser micromachining. By contrast, devices fabricated from the same batch using conventional photolithography and argon ion milling typically show higher $T_c$ values of 80–85 K, consistent with additional oxygen redistribution induced by prolonged thermal processing.

All samples exhibit a finite residual resistance below $T_c$ in three-terminal $R$–$T$ measurements, which is attributed mainly to contact resistance at the Bi-2212–electrode interface. As listed in Table I, the residual resistance at 20 K increases in the order Cu, Ag, and Cr. In particular, Sample C with Cr electrodes shows residual resistance more than two orders of magnitude larger than those of the other samples, which we attribute to partial oxidation of Cr during thermal evaporation, leading to reduced electrical conductivity of the electrode layer.

Sample B with Cu electrodes exhibits the lowest residual resistance among the three samples, despite previous reports of increased contact resistance for Cu on YBCO-based superconductors. In Bi-2212, the presence of $Bi_2O_2$ block layers may promote a more favorable metal–superconductor interface. Although Sample C showed pronounced self-Joule heating and reduced bias stability during current–voltage ($I$–$V$) measurements, terahertz emission was observed from all samples within the bath-temperature $T_b$ ranges listed in Table I. In the following sections, we therefore focus on Sample B with Cu electrodes, which exhibits the most stable electrical and emission characteristics.

Figure 2(a) shows the $I$–$V$ characteristics of Sample B measured at $T_b$'s between 10.0 and 40.0 K, after subtraction of the voltage drop associated with the residual resistance (4.6 Ω). With increasing $T_b$, the hysteresis in the $I$–$V$ characteristics is progressively reduced. At $T_b = 10.0$ K, partial overlap of switching events among IJJs appears in the voltage range of 2.1–3.1 V; however, once all IJJs enter the finite-voltage state, a monotonic return branch is observed upon decreasing the bias voltage.

A key feature of JPEs fabricated by UV laser micromachining is the collective switching of IJJs. The initial $I$–$V$ branch remains nearly linear up to the critical current, followed by an abrupt and simultaneous transition of most IJJs into the finite-voltage state. For example, at $T_b = 10.0$ K, the zero-voltage state is preserved up to 37.2 mA before a collective switch occurs. This behavior indicates a high degree of uniformity among the stacked IJJs along the $c$-axis. In contrast, mesas fabricated by conventional photolithography commonly exhibit trapezoidal cross-sections that lead to sequential switching due to junction-area variation.[31,32] Such geometrical nonuniformity has been shown to suppress spontaneous phase synchronization through distributed resonance frequencies.[33,34] The nearly vertical sidewalls achieved by UV laser micromachining therefore provide favorable conditions for phase-synchronized terahertz emission, consistent with the present observations.

The corresponding terahertz emission characteristics are shown in Fig. 2(b). Dome-shaped emission profiles are observed on the low-bias return branches, where self-heating effects are minimized, and the maximum emission intensity decreases monotonically with increasing $T_b$. Sawtooth-like modulations around $T_b = 30.0$ K are attributed to intermittent retrapping of a fraction of IJJs into the zero-voltage state, indicating that the emission intensity scales with the number of resistive junctions.

Representative emission spectra measured using a Martin–Puplett interferometer (MPI) are shown in the inset of Fig. 2(a), with oscillation peaks at $f_e = 0.285$ THz at $T_b = 10.0$ K and $f_e = 0.260$ THz at $T_b = 10.0$ K. As expected from the Josephson effect, the emission frequency scales linearly with the applied bias voltage, providing approximately 10% frequency tunability within the emission region. Because the MPI detection sensitivity depends on the polarization orientation relative to the wire-grid polarizer (WGP), spectral peak height does not directly represent emission power, motivating the polarization analysis discussed below.

From the Josephson relation $f_e = 2eV/(hN)$, the number $N$ of IJJs participating in emission, $N$, can be estimated.[12] Using the stable emission data at $T_b = 10.0$ K yields $N = 4160$, corresponding to a mesa height of approximately 6.4 µm based on the IJJ spacing of 1.54 nm in Bi-2212. This value agrees well with the machining depth estimated independently from SEM observations. At $T_b = 40.0$ K, $N = 2840$ is obtained, indicating that approximately 32% of IJJs are retrapped into the zero-voltage state and do not contribute to emission.

Comparison among devices with different electrode materials shows that Sample B with Cu electrodes exhibits relatively higher emission intensity, while Sample A with Ag electrodes maintains the widest temperature range for stable emission (see Fig. S2). Although quantitative comparison is limited by device-to-device variations, these results indicate that Cu electrodes provide emission performance comparable to, or slightly superior to, Ag electrodes, consistent with their lower residual resistance.

Polarization analysis was performed for the terahertz radiation emitted from Sample B to identify the dominant cavity resonance mode of the Bi-2212 mesa. The polarization dependence of the transmitted intensity was measured using a rotating WGP, with background suppression achieved by bias-modulation lock-in detection.

Figure 3 shows the polarization characteristics measured at $T_b = 10.0$ K, where the bolometer output is plotted as a function of the angle $\theta$ between the WGP transmission axis and the short-edge direction of the mesa. The data are well described by a sinusoidal dependence, from which the polarization ellipse parameters were extracted. Least-squares fitting yields an axis ratio $r_p = 3.55 \pm 0.63$ and an azimuth angle $\psi = -4.9° \pm 1.4°$, corresponding to an ellipticity of $1/r_p = 0.281 \pm 0.05$. The reconstructed polarization ellipse (Fig. 3, inset) has its major axis nearly parallel to the short edge of the mesa, indicating excitation of a cavity mode polarized along the width direction. A minor long-edge component is also present, similar to weak two-dimensional modes previously reported for mesas with small aspect ratios.[16]

The degree of polarization $P$ was evaluated from the polarization parameters obtained by least-squares fitting, yielding $P = 0.853 \pm 0.02$. This value is comparable to those reported for synchronized JPEs,[35,36] indicating that the radiation is dominated by a single coherent electromagnetic mode associated with phase-synchronized IJJs. Residual depolarization is attributed mainly to polarization mixing in the optical detection path and scattering from surface machining debris.

Based on the measured emission frequency $f_e$ and polarization anisotropy, the radiation is attributed to a geometric cavity resonance. For a rectangular mesa with width $w$ and length $L$, the cavity resonance frequency is given by[16]

$$f_{m,p}^c = \frac{c_0}{2n}\sqrt{\left(\frac{m}{w}\right)^2 + \left(\frac{p}{L}\right)^2},$$

where $c_0$ is the speed of light in vacuum and $n = 4.2$ is the refractive index of Bi-2212. For $m = 1$ and $p = 0$, the transverse magnetic TM(1,0) mode yields $f_{1,0}^c = 0.279$ THz, in close agreement with the experimentally observed emission frequency. This assignment is consistent with the polarization ellipse orientation. The next-higher mode, TM(1,1), would be expected to exhibit substantially weaker polarization anisotropy, contrary to the observed large axis ratio. The predominance of the TM(1,0) mode is likely promoted by geometrical asymmetry of the mesa and surrounding structures, including the Au-wire bonding configuration along the long-edge direction (Figs. 1(b-1) and 1(c)).

Although the laser beam spot size is close to the diffraction limit ($\approx 1.4$ μm), the resulting trench width and machining depth expand to approximately 10 μm and 6.4 μm, respectively, indicating that the characteristic length scale of the process is governed by thermal properties rather than optical resolution. We therefore attribute UV laser micromachining of Bi-2212 to anisotropic, thermally driven laser ablation. Strong absorption of 355-nm radiation generates rapid local heating, leading to melting and decomposition of the $CuO_2$ layer structure, followed by pressure-driven ejection and resolidification of molten material, as evidenced by the caldera-like debris morphology observed by SEM.

A key factor controlling the machining geometry is the highly anisotropic thermal conductivity of Bi-2212, which is approximately an order of magnitude larger in the $ab$-plane than along the $c$-axis.[37] Consequently, thermal diffusion during laser irradiation produces a laterally extended, disk-like temperature distribution instead of a cylindrical profile. The experimentally observed combination of a wide trench and a comparatively smaller machining depth is consistent with this anisotropic thermal diffusion picture. The reduced effective anisotropy in the final machining profile is likely influenced by transient UV absorption by redeposited debris and enhanced in-plane heat diffusion through the surface metal electrode layers.

In summary, we demonstrated terahertz JPEs based on cuprate high-$T_c$ superconducting Bi-2212 single crystals fabricated by direct UV laser micromachining. This maskless approach enables rapid patterning of single-crystal superconductors without conventional lithographic processing. Despite surface machining debris, the IJJs retain a high degree of uniformity, as evidenced by collective switching behavior and stable terahertz emission. Devices employing Cu electrodes exhibited emission performance comparable to Ag electrodes, identifying Cu as a low-cost and practical alternative. Spectroscopic and polarization analyses indicate that the emitted radiation is governed by the TM(1,0) cavity resonance mode. Structural analysis further shows that the machining width and depth are determined by the anisotropic thermal conductivity of Bi-2212 rather than by optical resolution, consistent with thermally dominated laser ablation.

Notably, although the strong thermal anisotropy of Bi-2212 tends to broaden the in-plane machining region, direct UV laser micromachining still achieves micrometer-scale patterning sufficient for functional device operation. This result demonstrates that the technique remains effective even in highly anisotropic cuprate superconductors. For other high-$T_c$ superconductors with weaker anisotropy, such as YBCO, further improvement in machining resolution and device performance is expected. The present approach therefore provides a practical and scalable fabrication strategy for high-$T_c$ superconducting electronics and terahertz devices,[38] including filters,[39] amplifiers,[40,41] detectors,[42] and superinductors.[43]

## ACKNOWLEDGEMENT

The authors thank A. Kitoh for technical support and Y. Saito and Y. Takano for experimental assistance and helpful discussions. This work was supported by JST PRESTO under Grant No. JPMJPR24F5, Japan.

## AUTHOR DECLARATIONS
### Conflict of Interest
The authors have no conflicts to disclose.
### Author Contributions
R. Yamaguchi and T. Sakurai performed device fabrication and experiments; they contributed equally to this work. K. Yamaki, A. Irie, J. Kato, and T. Nishio contributed to experimental support, data interpretation, and manuscript preparation. S. Ishida and H. Eisaki grew the Bi-2212 single crystals and contributed to discussions and manuscript preparation. M. Tsujimoto conceived and supervised the project, analyzed the data, and wrote the manuscript.

## DATA AVAILABILITY
The data that support the findings of this study are available from the corresponding author upon reasonable request.

# FIGURES AND TABLE

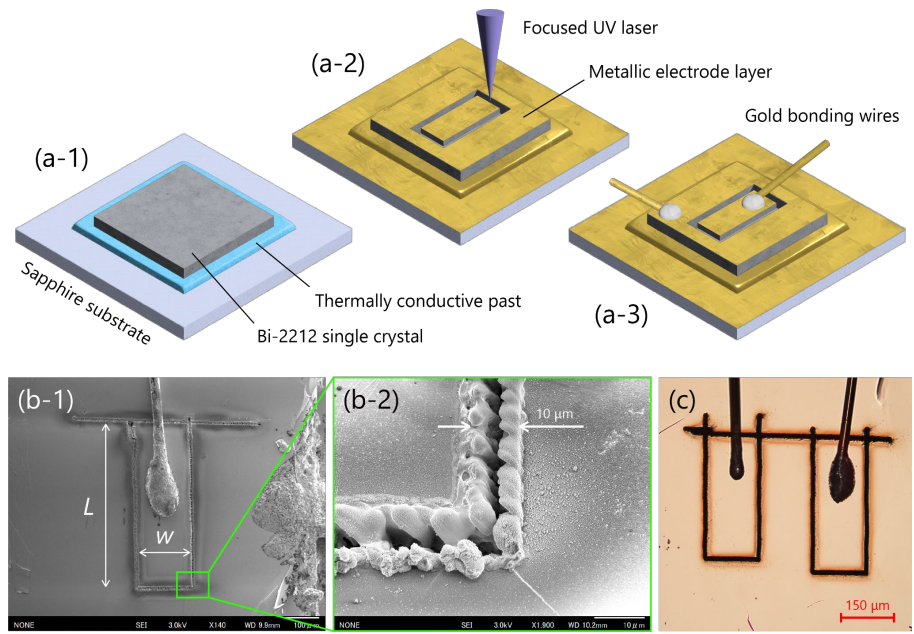

**FIG. 1.**
Fabrication process and morphology of Bi-2212 JPEs. (a-1) Bonding of a Bi-2212 single crystal onto a sapphire substrate, followed by cleavage and vacuum deposition of the electrodes with an Au capping layer. (a-2) Direct trenching of the Bi-2212 surface by UV laser micromachining. (a-3) Formation of the bias electrode by Au wire bonding onto the isolated mesa. (b-1) SEM image of a fabricated Bi-2212 mesa. (b-2) Magnified view of the region indicated in (b-1), showing a caldera-like morphology caused by redeposited machining debris. (c) Optical micrograph of a mesa array fabricated using the same process.

**TABLE I:**

Summary of the fabricated JPE samples, including the bottom electrode material, mesa dimensions (width $w$ and length $L$), $T_c$, $c$-axis resistivity $\rho_c$ at 20 K, and the $T_b$ range over which terahertz emission was observed.

| Sample | A | B | C |
| --- | --- | --- | --- |
| Electrode layers | Au (20 nm) / Ag (45 nm) | Au (20 nm) / Cu (80 nm) | Au (20 nm) / Cr (40 nm) |
| $w$ (μm) | 185 | 128 | 178 |
| $L$ (μm) | 382 | 374 | 422 |
| $T_c$ (K) | 77.0 | 75.8 | 75.6 |
| $\rho_c$ (Ω·cm) at 20.0 K | 2.3 | 2.0 | 242 |
| $T_b$ range for emission (K) | 10–50 | 10–40 | 10–40 |

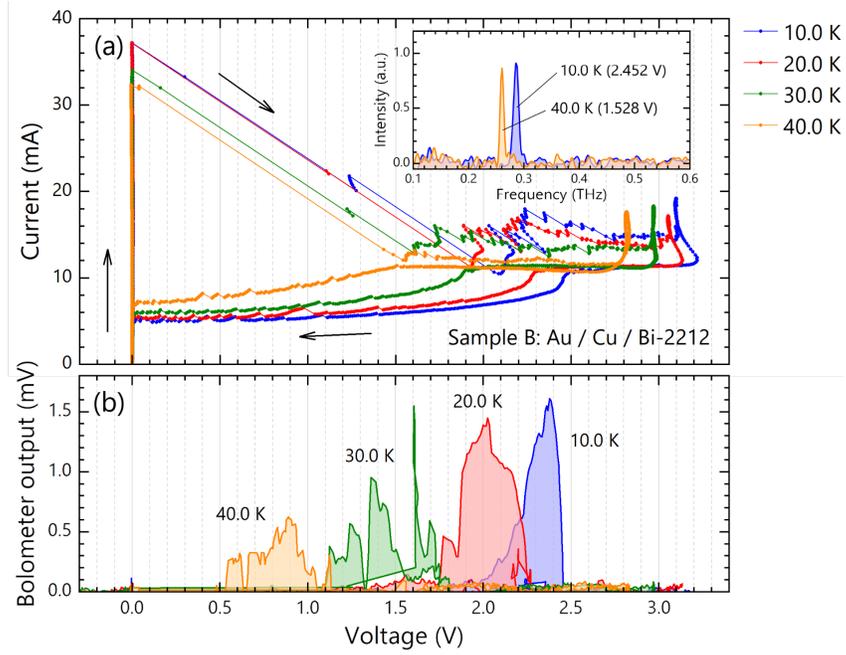

**FIG. 2.**
Electrical and terahertz emission characteristics of Sample B with Cu electrodes. (a) *I–V* characteristics measured at different $T_b$. Arrows indicate the bias sweep directions associated with intrinsic hysteresis of the IJJ stack. Inset: Emission spectra measured using a MPI at $T_b = 10.0$ K and 40.0 K. (b) Bolometer output as a function of bias voltage using the same voltage scale as in (a), with each curve corresponding to the indicated $T_b$.

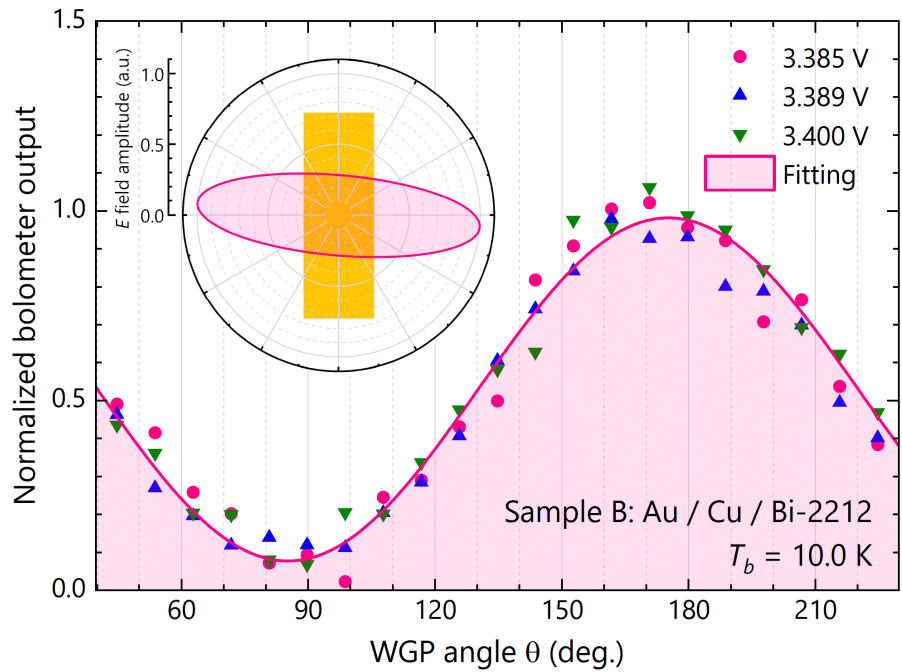

**FIG. 3.**
Polarization characteristics of terahertz radiation emitted from Sample B at $T_b = 10.0$ K. The horizontal axis $\theta$ denotes the angle between the transmission axis of the rotating WGP and the short-edge direction of the Bi-2212 mesa, and the vertical axis shows the normalized bolometer output. Symbols represent experimental data obtained at three bias points, and solid lines indicate least-squares fits. Inset: Polarization ellipse reconstructed from the fitted parameters; the rectangle indicates the mesa orientation.

# Supplemental Online Material

## 1. Temperature dependence of $c$-axis resistivity in JPEs with different electrode materials

Figure S1 shows the resistance–temperature ($R$–$T$) characteristics of Samples A (Ag electrode), B (Cu electrode), and C (Cr electrode), all fabricated under identical conditions using the direct UV laser micromachining process. The vertical axis represents the $c$-axis resistivity ($\rho_c$), normalized by the mesa width $w$, length $L$, and the machining depth of 6.4 μm estimated in the main text. All samples exhibit nearly identical superconducting transition temperatures ($T_c$), confirming that the bulk superconducting properties of the Bi-2212 single crystals are well preserved irrespective of the electrode material.

Below $T_c$, all samples show a finite residual resistivity, which arises primarily from the three-terminal measurement configuration and reflects the contact resistance at the interface between the Bi-2212 crystal and the metallic electrodes. Among the three samples, Sample C with Cr electrodes exhibits a $\rho_c$ value approximately two orders of magnitude larger than those of Samples A and B in the low-temperature regime, indicating significantly more insulating behavior. This enhanced resistivity is attributed to electrode oxidation and unfavorable interfacial properties. Because of the large residual resistance, Sample C exhibited pronounced self-Joule heating during current–voltage measurements, leading to unstable bias operation and reduced reproducibility of terahertz emission. In contrast, Samples A (Ag electrode) and B (Cu electrode) show relatively small residual resistivity at low temperatures, demonstrating good electrical contact at the Bi-2212–electrode interface.

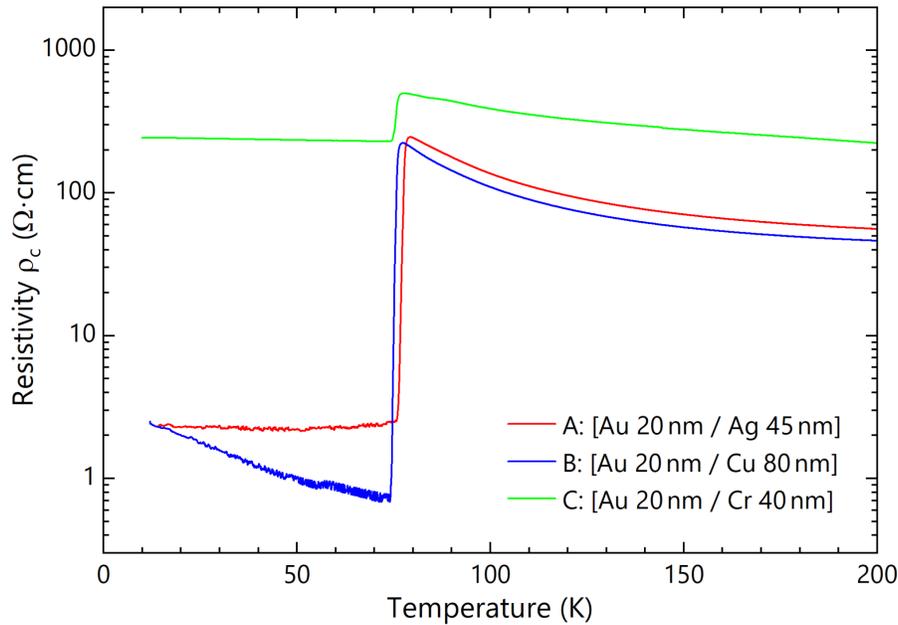

**FIG. S1.**
$R$–$T$ characteristics of Samples A–C listed in Table I. The vertical axis shows the $c$-axis resistivity $\rho_c$, normalized using the mesa width $w$, length $L$, and a thickness of 6.4 μm.

## 2. Comparison of terahertz emission power among JPEs with different electrode materials

To evaluate the influence of electrode material on terahertz emission performance, the emission characteristics of Samples A–C were compared using bolometric detection. Figure S2 plots the maximum bolometer output observed for each sample as a function of the bath temperature $T_b$. Sample B with Cu electrodes exhibits a relatively larger emission output than the other samples over the measured temperature range. In contrast, Sample A with Ag electrodes shows a slightly wider temperature window in which stable terahertz emission is observed.

For Sample C with Cr electrodes, the large residual resistance below $T_c$ results in significant self-Joule heating, which limits the stability of the bias point required for coherent terahertz emission. As a result, emission from Sample C is less stable and reproducible compared to Samples A and B. Considering unavoidable variations in mesa dimensions and intrinsic device-to-device differences, the present results do not allow a fully quantitative ranking of electrode materials. Nevertheless, Sample B with Cu electrodes maintains an emission temperature range comparable to that of Ag-electrode devices while exhibiting lower residual resistance, as discussed in the main text. These results suggest that Cu electrodes offer a practical and potentially advantageous alternative to Ag electrodes for JPE operation.

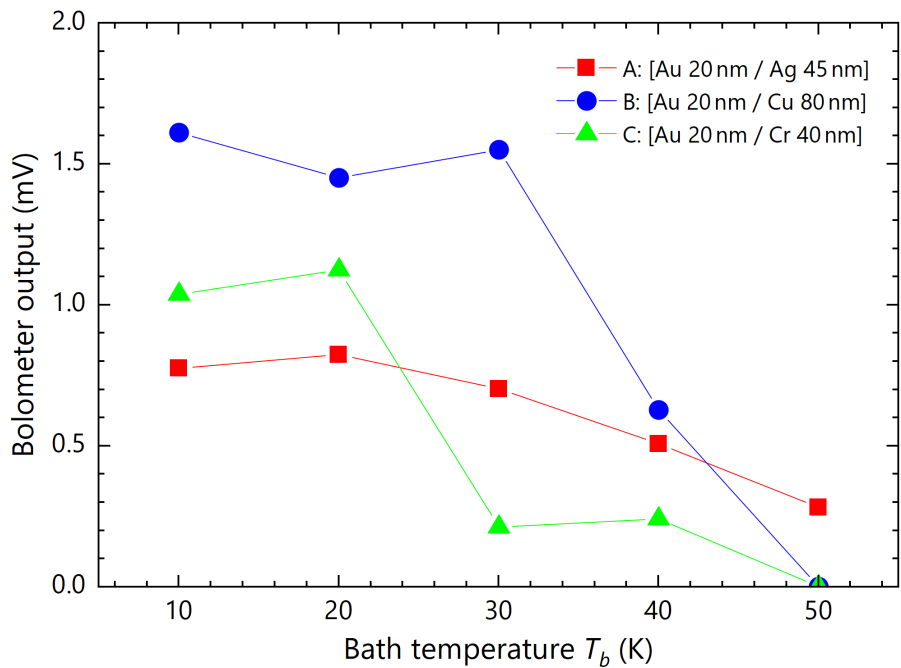

**FIG. S2.**
$T_b$ dependence of the terahertz emission output for Samples A, B, and C with different electrode materials.